\documentclass[A4,10pt]{iopart}
\usepackage{amsfonts}
\usepackage{amssymb}
\usepackage{geometry}
\usepackage{graphicx}
\usepackage{dsfont}
\usepackage{color}
\usepackage{hyperref}

\begin{document}

\title[Thermodynamics of gravitational clustering phenomena]{Thermodynamics of gravitational clustering phenomena:\\$N$-body self-gravitating gas on the sphere $\mathbb{S}^{3}\subset\mathbb{R}^{4}$}
\author{F. Tello-Ortiz and L. Velazquez}
\address{Departamento de F\'\i sica, Universidad Cat\'olica del Norte, Av. Angamos 0610, Antofagasta,
Chile.\newline
\newline
E-mail: ftello@ucn.cl; lvelazquez@ucn.cl}
\begin{abstract}
This work is devoted to the thermodynamics of gravitational clustering, a collective phenomenon with a great relevance in the $N$-body cosmological problem. We study a classical self-gravitating gas of identical non-relativistic particles defined on the sphere $\mathbb{S}^{3}\subset \mathbb{R}^{4}$ by considering gravitational interaction that corresponds to this geometric space. The analysis is performed within microcanonical description of an isolated Hamiltonian system by combining continuum approximation and steepest descend method. According to numerical solution of resulting equations, the gravitational clustering can be associated with two microcanonical phase transitions. A first phase transition with a continuous character is associated with breakdown of $SO(4)$ symmetry of this model. The second one is the gravitational collapse, whose continuous or discontinuous character crucially depends on the regularization of short-range divergence of gravitation potential. We also derive the thermodynamic limit of this model system, the astrophysical counterpart of Gibbs-Duhem relation, the order parameters that characterize its phase transitions and the equation of state. Other interesting behavior is the existence of states with negative heat capacities, which appear when the effects of gravitation turn dominant for energies sufficiently low. Finally, we comment the relevance of some of these results in the study of astrophysical and cosmological situations. Special interest deserves the gravitational modification of the equation of state due to the local inhomogeneities of matter distribution. Although this feature is systematically neglected in studies about Universe expansion, the same one is able to mimic an effect that is attributed to the dark energy: a negative pressure.

\end{abstract}

\tableofcontents

\section{Introduction}

Astrophysical structures observed in the Universe, including the Universe itself, pose challenging questions for the conventional thermodynamics and statistical mechanics. Gravitation is a long-range interaction that does not satisfy the \emph{stability and regularity properties} \cite{Reichl}-\cite{Gallavotti}, which are basic requirements for a self-consistent statistical mechanical description and the validity of theorems assuring the equivalence of the statistical ensembles in the thermodynamic limit. Considerable efforts have been devoted in the last decades for understanding thermo-statistical properties of long-range interacting systems, in particular, the astrophysical systems \cite{Gross}-\cite{Campa.2}. Although there still survive many open problems in this research field \cite{Vel.OpenAstro}, the understanding of the thermodynamic properties of astrophysical systems slowly emerges. Among their distinguishing features, one can mention the presence of negative heat capacities \cite{Eddington1}-\cite{Thirring}, instability processes like gravothermal collapse and evaporation disruption \cite{Antonov}-\cite{Vel.QEM2}, the non applicability of conventional thermodynamics limit and statistical ensembles \cite{Cerruti}-\cite{Vel.QEM2}, the incidence of long-range correlations and non-separability \cite{Gallavotti}-\cite{Labini}, and exotic collective phenomena like the microcanonical phase transitions \cite{Antoni}-\cite{Chavanis2004}.

This work is devoted to the thermo-statistics of gravitational clustering phenomenon. As expected, thermodynamical description provides a complementary view of this collective behavior beyond the standard picture obtained from $N$-body cosmological simulations. The motivation for this paper born from the reading of some precedent studies on this question available in the astrophysical literature \cite{SaslawHamilton}-\cite{Saslaw}. Surprisingly, most of these works start from the consideration of conventional ensembles of statistical mechanics such as canonical or gran-canonical ensembles. Typically, these studies start from the calculation of partition function:
\begin{equation}\label{ZZ}
Z_{N}(T,V)=\frac{1}{\Lambda^{N}N!}\int\exp\left[-\beta H_{N}(\mathbf{r},\mathbf{p})\right]d^{N}\mathbf{r}d^{N}\mathbf{p}
\end{equation}
associated with a $N$-body self-gravitating system of non-relativistic particles with Hamiltonian $H_{N}(\mathbf{r},\mathbf{p})$. Partition function depends on the system volume $V$ and its average temperature $T$, which enters into expression (\ref{ZZ}) via the control parameter $\beta=1/kT$. The constant $\Lambda$ normalizes the phase space volume cell. This type of statistical description, however, is not fully justified according to the present understanding about thermodynamics of systems with long-range interactions, astrophysical systems in particular. Originally, these theoretical frameworks were developed to deal with the macroscopic description of systems of everyday applications, which are composed of a very large number of constituents that interact among them by means of short-range interactions. For these systems, it is possible to speak about separability, additivity, extensivity and other macroscopic properties that support the licitness of conventional statistical ensembles. None of these properties, however, are met in the context of astrophysical systems. A more complete discussion about criticisms concerning the above thermo-statistical studies (and other related questions) were presented in a precedent paper \cite{Vel.OpenAstro}.

Our interest in this new contribution is to propose a very simple $N$-body thermo-statistical model that accounts for gravitational clustering phenomenon by including some basic ingredients with relevance in the cosmological scenario. Specifically, we shall analyze the thermo-statistics of a classical self-gravitating gas of identical non-relativistic particles defined on the sphere $\mathbb{S}^{3}\subset \mathbb{R}^{4}$ by considering gravitational interaction that corresponds to this particular geometry. As discussed elsewhere \cite{Bergstrom}, the curved geometry on the sphere $\mathbb{S}^{3}$ is one of the three possible solutions of Einstein field equations that is fully compatible with requirements of homogeneity and isotropy. Moreover, the closed character of this manifold avoids the incidence of evaporation, which also circumvents the usage of non trivial boundary conditions (e.g.: $N$-body dynamical simulations inside a cube of the real space $\mathbb{R}^{3}$ require the usage periodic boundary conditions, or alternatively, the consideration of impenetrable walls). Due to the closed character of this model system, the thermo-statistical description is performed considering the microcanonical ensemble, which describes an isolated system in thermodynamic equilibrium. This ensemble is the only conventional ensemble of statistical mechanics that enables the access to states with negative heat capacities \cite{Thirring}, which appear once gravitation becomes dominant in the dynamics and thermodynamics of this physical situation. Anticipating some of our results, gravitational clustering of this model can be associated with two microcanonical phase transitions: a continuous phase transition associated with breakdown of $SO(4)$ symmetry (and the uniform distribution of the matter on the sphere $\mathbb{S}^{3}$), and a second one related to gravitational collapse that leads to the formation of core-halo structures, which is also accompanied by the presence of states with negative heat capacities. The character continuous or discontinuous of this second phase transition crucially depends on the details of regularization scheme of short-range divergence of gravitation.

\section{Methodology}

\subsection{Green's solution on the sphere $\mathbb{S}^{3}$ and the model system}

In the non-relativistic limit, gravitation is described in terms of gravitational potential $\Phi(\mathbf{x})$ through Poisson equation:
\begin{equation}\label{poisson}
\Delta \Phi(\mathbf{x})= 4 \pi G \rho(\mathbf{x}),
\end{equation}
where $\rho(\mathbf{x})$ is the particles density and $\Delta$ is Laplace-Beltrami operator:
\begin{equation}
\Delta \Phi(\mathbf{x})=\frac{1}{\sqrt{\vert g_{\mu\nu}(\mathbf{x})\vert }}\frac{\partial}{\partial x^{\alpha}}\left[\sqrt{\vert g_{\mu\nu}(\mathbf{x})\vert }g^{\alpha\beta}(\mathbf{x})
\frac{\partial}{\partial x^{\beta}}\Phi(\mathbf{x})\right].
\end{equation}
Here, $g_{\alpha\beta}(\mathbf{x})$ is the spatial part of the metric tensor. Let us consider the space geometry corresponding to a three-dimensional sphere $\mathbb{S}^{3}\subset \mathbb{R}^{4}$ of radius $R$:
\begin{equation} \label{distance}
d\ell^2=g_{\alpha\beta}(\mathbf{x})dx^{\alpha}dx^{\beta}=R^2\left[d\phi^2 + \sin^2 \phi \left( d\varphi^2 + \sin^2\varphi d\theta^2 \right)\right],
\end{equation}
with  axial $\theta$ and azimuthal angles $\phi$ and $\varphi$.

\begin{figure}
  \centering
  \includegraphics[width=4.0in]{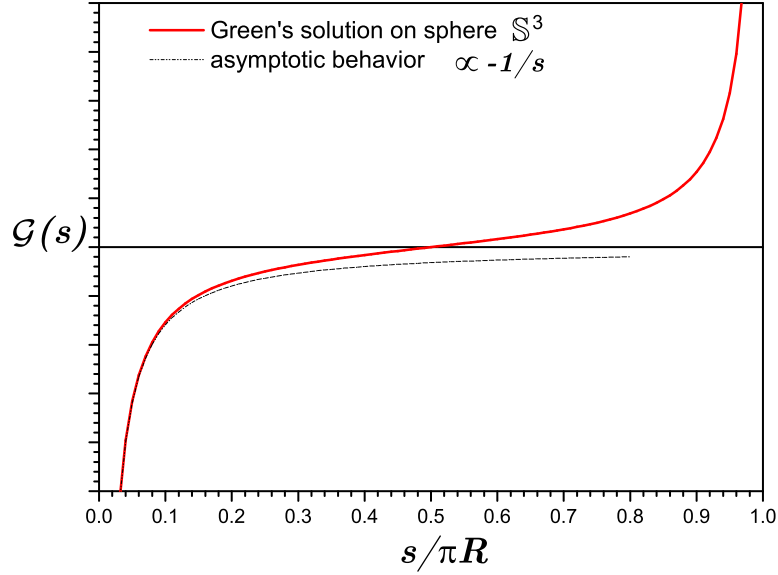}\\
  \caption{Dependence of Green's solution (\ref{cot}) on the separation $s$ (solid line) and its asymptotic dependence $-1/s$ for small distances.}\label{potencial.eps}
\end{figure}

Green's solution of problem (\ref{poisson}) can be expressed as follows:
\begin{equation}\label{Green}
\Phi(\mathbf{x})=\int_{\mathbb{S}^{3}} \mathcal{G}(\mathbf{x},\mathbf{x}')\rho(\mathbf{x}')d\mu(\mathbf{x}')
\end{equation}
with volume element:
\begin{equation}\label{dmu}
d\mu(\mathbf{x})=\sqrt{\vert g_{\mu\nu}(\mathbf{x})\vert }d^{3}x=R^{3}\sin^{2}{\phi}\sin{\varphi}d\phi d\varphi d\theta,
\end{equation}
where $\mathcal{G}(\mathbf{x},\mathbf{x}')$ is the gravitational potential for a particle of unitary mass located at the point $x^{\prime}$. As a consequence of underlying symmetry of sphere $\mathbb{S}^{3}$ under rotation group $SO(4)$, Green's function $\mathcal{G}(\mathbf{x},\mathbf{x'})$ depends on the separation distance $s=s(\mathbf{x},\mathbf{x'})$ between the points $(\mathbf{x},\mathbf{x^{\prime}})$:
\begin{equation}
\mathcal{G}(\mathbf{x},\mathbf{x'})\equiv \mathcal{G}[s(\mathbf{x},\mathbf{x^{\prime}})],
\end{equation}
where $s(\mathbf{x},\mathbf{x^{\prime}})$ is evaluated as follows:
\begin{eqnarray}
s(\mathbf{x},\mathbf{x^{\prime}})=R \cos^{-1}\left\{\cos\phi \cos\phi' + \sin\phi \sin\phi' \left[\cos\varphi \cos\varphi'+\right.\right.\\
\left.\left.+ \sin\varphi \sin\varphi' \cos(\theta-\theta') \right]\right\}\nonumber
\end{eqnarray}
for two points with spherical coordinates $\mathbf{x}=(\phi,\varphi,\theta)$ and $\mathbf{x'}=(\phi',\varphi',\theta')$. For the sake of convenience, let us regard the reference point $\mathbf{x}^{\prime}$ at north pole of the sphere, where the azimuthal angle $\phi^{\prime}=0$. This restriction allows us to express the separation distance as $s\equiv R \phi$. Performing volume integration of Poisson's equation (\ref{poisson}) inside a two-sphere $\mathbb{S}^{2}\subset \mathbb{S}^{3}$ with radius $s$ that is centered at $x^{\prime}$, one obtains the following relation after applying divergence theorem:
\begin{equation}\label{gradiente}
\frac{1}{R} \frac{\partial \mathcal{G}(\phi)}{\partial{\phi}}4\pi R^{2} \sin^{2}\phi=4\pi G.
\end{equation}
Consequently, Green's function:
\begin{equation}\label{cot}
\mathcal{G}(s)=-\frac{G}{R}\cot(s/R)
\end{equation}
is obtained by integrating (\ref{gradiente}) over azimuthal angle $\phi$ and re-writing the result in terms of separation distance. It is noteworthy that function (\ref{cot}) diverges towards $-\infty$ when separation distance $s\rightarrow 0$, while it diverges towards $+\infty$ when $s\rightarrow \pi R$. Consequently, Green's function (\ref{cot}) actually corresponds to the Poisson's problem:
\begin{equation}\label{poisson2}
\Delta \mathcal{G}(\mathbf{x},\mathbf{x}')=4\pi G \frac{1}{\sqrt{\vert g_{\mu\nu}(\mathbf{x})\vert }}\left[\delta(\mathbf{x}-\mathbf{x}')-\delta(\mathbf{x}-\mathbf{x}^{*})\right],
\end{equation}
where $\mathbf{x}^{*}$ denotes a point on the sphere $\mathbb{S}^{3}$ that is diametrically opposed to the point $\mathbf{x}'$, that is, $\mathbf{x}$ and $\mathbf{x}'$ are antipodal points of the sphere $\mathbb{S}^{3}$. As expected, one obtains the well-known Newtonian limit $\mathcal{G}(s)\simeq -1/s$ for small distances $s\ll R$. The mathematical behavior of Green's solution (\ref{cot}) is shown in figure \ref{potencial.eps}. It is worth mentioning that the Green solution (\ref{cot}) is the difference of the Green functions with singularity at north $\mathcal{G}^{\left( +\right) }\left( s\right) $ and the one with singularity at south $\mathcal{G}^{\left( -\right) }\left( s\right)$:
\begin{equation}\label{true.Green}
\mathcal{G}\left( s\right) =\mathcal{G}^{\left( +\right) }\left( s\right) -\mathcal{G}^{\left( -\right) }\left( s\right).
\end{equation}
Topologically speaking, it is unavoidable that the existence of a singularity of lines of forces of gravitation potential at a given point on the sphere $\mathbb{S}^{3}$ (e.g.: the north pole) will force the existence of the anti-singularity at its antipodal point (e.g.: the south pole). In our opinion, any prescription to separate such north and south contributions is always a matter of convention. For example, one could employ the following prescriptions:
\begin{equation}
\mathcal{G} ^{\left( +\right) }\left( s\right) =-\frac{G}{R}\cot \left( \frac{s}{R}
\right) \Theta \left( \frac{\pi }{2}-\frac{s}{R}\right) \mbox{ and }
\mathcal{G}^{\left( -\right) }\left( s\right) =\mathcal{G}^{\left( +\right) }\left( \pi R-s\right),
\end{equation}
which respectively correspond to a point particle with unitary mass located at the north (and the south) pole plus a homogeneous distribution of a unitary \emph{negative mass} on the sphere $\mathbb{S}^{2}_{r}\subset \mathbb{S}^{3}$ with maximum radius $r=\pi R/2$ (the equatorial sphere). Other possible and convenient alternative is the consideration of a point particle of unitary mass at a given point embedded into a uniform sea background with a negative total unitary mass over the whole sphere $\mathbb{S}^{3}$. However, none of the above alternatives is able to meet a basic property of Green solution in $\mathbb{R}^{3}$: to be the solution of Poisson problem for an isolated point particle located at a given point of the ambient space. Although it would be interesting to check the influence of different possible solutions of Green function, we shall restrict in this work to deal with the solution (\ref{true.Green}).

The self-gravitating gas on the sphere $\mathbb{S}^{3}$ can be described in non-relativistic limit by the Hamiltonian:
\begin{equation}\label{hamilton}
H_{N}(\mathbf{x},\mathbf{p})=K_{N}(\mathbf{x},\mathbf{p})+W_{N}(\mathbf{x})
=\sum_{i} \frac{1}{2m_{i}}\mathbf{p}^{2}_{i} + \sum_{i < j} w(\mathbf{x}_{i},\mathbf{x}_{j}).
\end{equation}
Here, $K_{N}(\mathbf{x},\mathbf{p})$ represents the total kinetic energy, where $\mathbf{p}^{2}\equiv g^{\alpha\beta}(\mathbf{x})p_{\alpha}p_{\beta}$ is the square norm of momentum {${\mathbf{p}}$} with covariant components $\left\{p_{\alpha}\right\}$. Moreover, $(\mathbf{x}_{i},\mathbf{p}_{i})$ are the position and the momentum of $i$-th particle with mass $m_{i}$. Latin indexes $i$ and $j$ run over all $N$ constituting particles of this system. On the other hand, $W_{N}(\mathbf{x})$ represents the total potential energy, where $w(\mathbf{x}_{i},\mathbf{x}_{j})$ denotes the interaction potential associated with two particles located at positions  $\mathbf{x}_{i}$ and $\mathbf{x}_{j}$:
\begin{equation}\label{interact.w}
w(\mathbf{x}_{i},\mathbf{x}_{j})=m_{i}m_{j}\mathcal{G}(\mathbf{x}_{i},\mathbf{x}_{j})
=-\frac{Gm_{i}m_{j}}{R}\cot\left[\frac{s(\mathbf{x}_{i},\mathbf{x}_{j})}{R}\right].
\end{equation}
Although we are restricting to the non-relativistic limit, this system exhibits an explicit invariance under coordinate representations of the sphere $\mathbb{S}^{3}$. Notice that gravitational potential $\Phi(\mathbf{x})$, the particles density $\rho(\mathbf{x})$ and the Hamiltonian behave here as scalar functions. For the sake of simplicity, we shall hereinafter assume that all system particles are identical with mass $m_{i}=m$.

\subsection{Microcanonical thermo-statistics in the continuum approximation}

The thermo-statistical study of the closed Hamiltonian system (\ref{hamilton}) can be performed in the framework of microcanonical ensemble. The number of microstates is expressed as follows:
\begin{equation}\label{NE}
\Omega_{N}(U)=\frac{1}{N!} \int \Theta\left[U - H_{N}(\mathbf{x},\mathbf{p}) \right] \prod_{i}\frac{d^{3}\mathbf{x}_{i} d^{3}\mathbf{p}_{i}}{(2\pi\hbar)^{3}},
\end{equation}
where $\Theta(x)$ is the Heaviside step function. The integration over momenta yields:
\begin{equation}\label{NE1}
\Omega_{N}(U)=\mathcal{C}\int \left[U-W_{N}(\mathbf{x})\right]^{\frac{3}{2}N}\prod^{N}_{i=1}d\mu(\mathbf{x}_{i}),
\end{equation}
where $\mathcal{C}$ is the integration factor:
\begin{equation}
\mathcal{C}=\frac{1}{N! \Gamma(\frac{3}{2}N + 1)}\left(\frac{ m }{2\pi\hbar^{2}}\right)^{\frac{3}{2}N}.
\end{equation}
Direct integration over positions cannot be performed, but an analytical treatment is still possible invoking the \emph{continuum approximation}, which is valid for a large number of particles $N$. This approximation was discussed in details by Padmanabhan in Ref.\cite{Padmanabhan1}, so that, let us restrict here to present the final results obtained from this procedure. A distinguishing feature of the present development is the preservation of the invariance of all functional dependencies under local diffeomorphims of the background geometric space. The number of microstates (\ref{NE1}) can be expressed in term of the following functional integral:
\begin{equation}\label{W2}
\Omega_{N}(U)=\int D \bar \varrho(\mathbf{x})\delta\{N-N[\varrho]\}\exp\{S[\varrho]\}.
\end{equation}
Here, $S[\varrho]$ denotes the entropy functional:
\begin{eqnarray}\label{entropia}
S[\varrho]&=&\frac{3}{2}N\log\left(\frac{m}{3 \pi\hbar^{2}N} \left\{U-W[\varrho]\right\} \right)+\\
 &+& N\log \left( \frac{e\mathcal{A}}{N}\right)+ N\int_{\mathbb{S}^{3}}\varrho(\mathbf{x})\log[\bar\varrho(\mathbf{x})]d\mu(\mathbf{x}),\nonumber
\end{eqnarray}
while $W[\varrho]$ and $N[\varrho]$ represent the total potential energy and the number of particles functionals in the continuum approximation:
\begin{eqnarray}\label{VV}
W[\varrho]&=&\frac{1}{2}N^{2}\int_{\mathbb{S}^{3}}\int_{\mathbb{S}^{3}}w(\mathbf{x},\mathbf{x}')
\varrho(\mathbf{x})d\mu(\mathbf{x})\varrho(\mathbf{x}')d\mu(\mathbf{x}'),\\
N[\varrho]&=&N\int_{\mathbb{S}^{3}}\varrho(\mathbf{x})d\mu(\mathbf{x}).
\end{eqnarray}
All these functionals are expressed in terms of the probability density $\varrho(\mathbf{x})$, which determines the occupation probability that a particle is located at the position $\mathbf{x}\in \mathbb{S}^{3}$:
\begin{equation}
dp(\mathbf{x})=\varrho(\mathbf{x})d\mu(\mathbf{x}).
\end{equation}
Moreover, $\mathcal{A}=2{\pi}^{2} R^{3}$ is the hiper-surface area (volume) of the sphere $\mathbb{S}^{3}$, while $\bar{\varrho}(\mathbf{x})$ is the dimensionless probability density $\bar{\varrho}(\mathbf{x})=\varrho(\mathbf{x})\mathcal{A}$. The analytical treatment of the functional integral (\ref{W2}) is a hard task. Nevertheless, one can obtain a good estimation for the entropy $S$ applying the \emph{steepest descent method} as follows:
\begin{equation}
S=k\log \Omega_{N}(U)\simeq max_{\varrho}min_{\lambda} S[\varrho] + \lambda \{ N-N[\varrho]\},
\end{equation}
where $\lambda$ is the Lagrange's multiplier. Since the probability density $\varrho(\mathbf{x})$ is a scalar function, we need to employ the following rule for the functional differentiation:
\begin{equation}
\frac{\delta \varrho(\mathbf{x})}{\delta \varrho(\mathbf{y})}=\frac{1}{\sqrt{\left|g_{\alpha\beta}(\mathbf{x})\right|}}\delta\left[\mathbf{x}-\mathbf{y}\right]
\end{equation}
in order to preserve the underlying covariance of the thermo-statistical description. This procedure yields the most likely configuration for probability density:
\begin{equation}\label{varrho}
\varrho(\mathbf{x})=C \exp[-\beta m\Phi(\mathbf{x})]
\end{equation}
with the self-consistent condition for the gravitational potential $\Phi(\mathbf{x})$ at the position $\mathbf{x}$:
\begin{equation}\label{cond.auto}
m\Phi(\mathbf{x})=N\int_{\mathbb{S}^{3}}w(\mathbf{x},\mathbf{x}^{\prime})\varrho(\mathbf{x}^{\prime})d\mu(\mathbf{x}^{\prime}).
\end{equation}
Additionally, this solution must satisfy the following constraints:
\begin{equation}\label{cond.rest}
N[\varrho]=N\mbox{ and }\beta=\frac{3}{2}\frac{N}{U-W[\varrho]},
\end{equation}
which are respectively associated with conservation of the particles number and the total energy, with $\beta=1/kT$ being the inverse temperature parameter. Moreover, the gravitational potential energy $W\left[\varrho\right]$ is evaluated through the formula:
\begin{equation}
W\left[\varrho\right]=\frac{1}{2}N\int_{\mathbb{S}^{3}}m\Phi(\mathbf{x})\varrho(\mathbf{x})d\mu(\mathbf{x}).
\end{equation}
Expression (\ref{varrho}) is the known Boltzmann distribution. At first glance, one might think that this result suggests the equivalence between canonical and microcanonical descriptions. However, this is a wrong idea. The energy $U$ is the control parameter of the present equilibrium situation, while the inverse temperature $\beta$ is merely a thermodynamic quantity that depends on the energy. We shall verify that relationship between these quantities, the called microcanonical caloric curve, exhibits nontrivial mathematical behaviors that discard the equivalence of ensembles.

\subsection{Computational study of configurations with symmetry $SO(3)$}\label{comput.proc}
The natural analytical treatment for the self-consistent nonlinear problem (\ref{varrho})-(\ref{cond.rest}) involves the generalized spherical harmonics on the sphere $\mathbb{S}^{3}$ \cite{Abramowitz}. However, the nonlinear character of this problem unavoidably demands a numerical treatment (see an analogous development in Ref.\cite{Votyakov}). For the sake of convenience, let us attempt to obtain a numerical solution employing \emph{finite elements method}. The occurrence of gravitational clustering will involve a breakdown of the underlying $SO(4)$ symmetry of the Hamiltonian (\ref{hamilton}). After this process, configurations with lower symmetry are still possible, e.g.: the symmetry $SO(3)$ respect to a certain point of the sphere $\mathbb{S}^{3}$. Let us develop our numerical analysis to derive those solutions with this type of residual symmetry. Without lost of generality, let us suppose that the point of symmetry is the north pole of the sphere $\mathbb{S}^{3}$. Notice that the residual symmetry $SO(3)$ is fully analogous to configurations with spherical symmetry in the corresponding problem in the Euclidean space $\mathbb{R}^{3}$, where the particles distribution depends on the radial distance with respect to the point of symmetry. According to this previous considerations, the particles distribution $\varrho(\mathbf{x})$ depends of the azimuthal angle $\phi$ only:
\begin{equation}\label{azimuthal.sym}
\varrho(\mathbf{x})\equiv \varrho(\phi),
\end{equation}
which also includes the homogeneous distribution $\varrho(\phi)=1/\mathcal{A}=const$ as a particular solution with symmetry $SO(4)$. Integrating expression (\ref{cond.auto}) over the angles $({\varphi}',{\theta}')$, it can be shown that the gravitational potential depends of the azimuthal angle $\phi$ as follows:
\begin{equation}\label{oo}
m\Phi(\mathbf{x})\equiv m\Phi(\phi)=N\int^{\pi}_{0} \varrho(\mathbf{\phi}^{\prime})v(\phi,\phi^{\prime})d\mu(\phi^{\prime}),
\end{equation}
where $d\mu(\phi)=4\pi R^{3}\sin^{2}\phi d\phi$ is the volume element of the spherical shell with azimuthal angle $\phi \in [\phi,\phi +d\phi]$. Moreover, the function $v(\phi,\phi^{\prime})$:
\begin{equation}\label{pot.efect}
v(\phi,\phi^{\prime})=v(\phi^{\prime},\phi)=\frac{Gm^{2}}{2R}\frac{\vert \sin(\phi-\phi^{\prime})\vert -\vert \sin(\phi+\phi^{\prime})\vert }{\sin\phi\sin\phi^{\prime}}
\end{equation}
is the effective interaction potential between two concentric spheres $\mathbb{S}^{2}\in \mathbb{S}^{3}$ with angles $\phi'$ and $\phi$ that are centered at the north pole of the sphere $\mathbb{S}^{3}$.

The application of finite element scheme starts partitioning the volume of the sphere $\mathbb{S}^{3}$ into a set of $P$ non-overlapping concentric shells with the same volume that are centered at the north pole of the sphere $\mathbb{S}^{3}$. Let us also assume that the particles density $\varrho(\mathbf{x})$ is uniformly distributed inside each shell. Formally, this is equivalent to suppose the existence of a finite volume for particles $\delta\mu$ and the upper bound for particles density (or occupation probability). Considering the extreme case where all particles are located inside a certain partition, one obtains the relation $\delta\mu=\mathcal{A}/NP\rightarrow \max\varrho(\mathbf{x})\leq 1/\delta\mu N=P/\mathcal{A}$.
The volume inside the sphere $\mathbb{S}^{2}$ of radius $s=R \phi$ centered at north pole is given by:
\begin{equation}
\mu(\phi)=2\pi R^{3}\left[\phi-\frac{1}{2}\sin(2\phi)\right].
\end{equation}
This last expression is employed here to obtain the reference azimuthal angle $\phi_{i}$ of the $i$-th spherical shell:
\begin{equation}
\frac{1}{P}\left(i-\frac{1}{2}\right)=\frac{\mu(\phi_{i})}{2\pi^{2}R^{3}},
\end{equation}
where the index $i$ run over for all possible shells of the partition. The azimuthal angles $\left\{\phi_{1},\phi_{2},\ldots,\phi_{P}\right\}$ are employed to rephrase the integral (\ref{oo}) into the discrete form:
\begin{equation}\label{Phi.matrix}
m \Phi_{i}=N\sum^{P}_{j=1}v(\phi_{i},\phi_{j})p_{j},
\end{equation}
while the total interaction energy (\ref{VV}) can be expressed as:
\begin{equation}\label{V.mat}
W=\frac{1}{2}N^{2}\sum^{P}_{i=1}m\Phi_{i}p_{i}.
\end{equation}
We have introduced here the occupation probability of the $i$-th shell, $p_{i}=\varrho_{i}\mu_{i}$, which is calculated as follows:
\begin{equation}\label{disc.pp}
p_{i}=\frac{1}{Z}\exp\left[-\beta m\Phi_{i}\right]\mbox{ with }Z=\sum_{i}\exp\left[-\beta m\Phi_{i}\right],
\end{equation}
when $P$ is large enough in accordance with expression (\ref{varrho}). The inverse temperature $\beta$ is given by:
\begin{equation}\label{beta.disc}
\beta=\frac{3}{2}\frac{N}{U-W}.
\end{equation}

Re-interpreting the occupation probabilities as the components of a vector $p=\left[p_{1},p_{2},\ldots,p_{P}\right]$, as well as the gravitational potential $\Phi=\left[\Phi_{1},\Phi_{2},\ldots,\Phi_{P}\right]$, the expression (\ref{Phi.matrix}) represents a matrix multiplication, while the relation (\ref{V.mat}) as a scalar product. This algebraic interpretation allows the implementation of parallel execution algorithms in FORTRAN $90$ programming. The solution of the self consistent problem (\ref{Phi.matrix})-(\ref{beta.disc}) is obtained using the method of successive iteration. The scheme implemented in this study is the following:
\begin{description}
  \item[i)] The $s$-th estimate of the occupation probability vector $p^{(s)}$ is employed to calculate the gravitational potential vectors $\Phi^{(s)}$ according to expression (\ref{Phi.matrix}).
  \item[ii)] The vectors $p^{(s)}$ and $\Phi^{(s)}$ are employed to calculate the total interaction energy $W^{(s)}$ and the inverse temperature $\beta^{(s)}$ via equations (\ref{V.mat}) and (\ref{beta.disc}).
  \item[iii)] The quantities $\beta^{(s)}$ and $\Phi^{(s)}$ are considered to obtain a \emph{tentative approximation} of the occupation probability vector $\tilde{p}^{(s+1)}$ via relation (\ref{disc.pp}).
  \item[iv)] The occupation probability vector for the next iteration $p^{(s+1)}$ is obtained from the average of its previous value $p^{(s)}$ and the tentative $\tilde{p}^{(s)}$ according the relation:
  \begin{equation}
    p^{(s+1)}=\sigma p^{(s)}+\left(1-\sigma\right)\tilde{p}^{(s+1)},
  \end{equation}
  being $\sigma$ some weighting factor, $0\leq\sigma<1$.
  \item[v)] The previous steps are repeated until the convergence error:
  \begin{equation}
  \epsilon[p]=\sqrt{\frac{1}{P}\sum^{P}_{i=1}\left (\tilde{p}^{(s+1)}-p^{(s)}\right)^{2}}< \epsilon
  \end{equation}
  reaches a prefixed accuracy $\epsilon$.
\end{description}

We have considered in our analysis two different initial conditions for the occupation probability vector. The first initial condition is the uniform distribution, $\forall i$ $p_{i}=1/P$ , which corresponds to the high energy limit. For the second one, the vector components $p_{i}$ are only non vanishing for the shell that contains the north pole of the sphere $\mathbb{S}^{3}$, $p_{i}=\delta_{1i}$, which corresponds to the system configuration with lowest energy. Once reached the convergence for a certain value of energy $U$, the thermodynamic observable are stored, and then, the resulting occupation probabilities vector $p$ is considered as the initial condition for a process of successive iterations for a new energy value $U+\delta U$, where $\delta U$ is a small energy step.

\subsection{Order parameters for the gravitational clustering}

As discussed elsewhere \cite{Reichl}, an \emph{order parameter} is an observable that characterizes the degree of ordering that acquires a system when the same one undergoes a phase transition. Typically, an order parameter exhibits a zero value in one phase (in general over the critical point) and non zero in other phase. In particular, gravitational clustering of the present model can be characterized by two different order parameters. Let us consider the vectorial quantity $\mathbf{M}\in \mathbb{R}^{4}$:
\begin{equation}\label{MM}
\mathbf{M}=\frac{1}{N}\sum^{N}_{i=1}\mathbf{S}_{i},
\end{equation}
defined from the positions of particles $\mathbf{x}_{i}$ on the sphere $\mathbb{S}^{3}$ in Cartesian coordinates of the Euclidean real space $\mathbb{R}^{4}$, where $\mathbf{S}_{i}=\mathbf{x}_{i}/R$. Formally, this vector represents the position of the center of mass of the self-gravitating gas in the ambient space $\mathbb{R}^{4}$, which is expressed in units of the radius $R$ of sphere $\mathbb{S}^{3}$. Since each vector $\mathbf{S}_{i}$ is unitary, $\left\vert \mathbf{S}_{i}\right\vert ^{2}\equiv 1$, the same one can also be interpreted as a four-vector of \emph{spin}. This consideration enables a direct analogy between self-gravitating gas on the sphere $\mathbb{S}^{3}$ and a \emph{ferromagnetic system}, where the vector $\mathbf{M}$ represents magnetization per particle. The magnetization vector $\mathbf{M}$ vanishes if the self-gravitating gas is homogeneously distributed on the sphere $\mathbb{S}^{3}$. Otherwise, the non-vanishing of this vector implies an non-homogenous distribution. The first order parameter is the modulus of magnetization per particle $M=\left\vert \mathbf{M}\right\vert $. In the framework of the continuum approximation, the magnetization vector (\ref{MM}) is rewritten in terms of the occupation density $\varrho(\mathbf{x})$ as follows:
\begin{equation}
\mathbf{M}=\int_{\mathbb{S}^{3}}\mathbf{S}(\mathbf{x})\varrho(\mathbf{x})d\mu(\mathbf{x}).
\end{equation}
Our computational approach was restricted to those distributions with $SO(3)$ symmetry with respect to north pole of the sphere $\mathbb{S}^{3}$. Accordingly, the only non-vanishing component of the magnetization vector $\mathbf{M}=(M_{1},M_{2},M_{3},M_{4})$ is the component $M_{1}$:
\begin{equation}
M_{1}=\int^{\pi}_{0}\cos\phi\,\varrho(\phi)d\mu(\phi),
\end{equation}
whose expression in finite elements reads as follows:
\begin{equation}
M_{1}=\sum_{i}\cos\phi_{i}\,p_{i}.
\end{equation}

A second order parameter that characterizes the occurrence of gravitational clustering is the total gravitational energy $W$, or more precisely, its absolute value $\vert W\vert $. According to Green function (\ref{cot}), this function exhibits both positive and negative values. In fact, the total gravitational energy is skew-symmetric under the inversion $s\rightarrow s^{*}=\pi R-s$. The total gravitational energy $W$ identically vanishes for a uniform distribution of particles on the sphere $\mathbb{S}^{3}$, whereas its non-vanishing indicates an inhomogeneous distribution. The explicit presence of this order parameter in the state equation (\ref{state.equation}) indicates when the self-gravitating gas starts to exhibit a non ideal behavior.

\section{Results and discussions}

\subsection{Thermodynamic limit and astrophysical Gibbs-Duhem relation}
For the sake of convenience, let us express the thermodynamic quantities into dimensionless units. The total gravitational energy $W$ as well as the total energy $U$ can be expressed in units of the characteristic energy $U_{c}=GM^{2}/R$, with $M=Nm$ being the total system mass. One can introduce the dimensionless energy $u=UR/GM^{2}$ and the dimensionless temperature $t=kTR/GMm$. According to these units, the system entropy (\ref{entropia}) can be written in the form:
\begin{equation}\label{Sreducida}
S=Nkf(u)+\frac{1}{2}Nk\log\left[\frac{2e^{2}}{27\pi}\frac{N \mathcal{A}}{\ell^{3}_{G}}\right],
\end{equation}
where $\ell_{G}=\hbar^{2}/Gm^{3}$ is the gravitational analogous of Bohr radius $a_{B}=\hbar/e^{2}m$. For extensive systems of everyday applications of thermodynamics, the total system volume $V$ scales proportional to the number of particles $N$, so that, the particles density remains constant. This scaling behavior ensures that both the energy $U$ and the entropy $S$ represent extensive quantities, which is associated with the thermodynamic limit:
\begin{equation}
N\rightarrow+\infty:\frac{N}{V}=const, \frac{U}{N}=const.
\end{equation}
This extensive thermodynamic limit does not apply for astrophysics systems because of the long-range character of gravitation. Expression (\ref{Sreducida}) evidences that the entropy $S$ remains extensive whenever one imposes the following thermodynamic limit:
\begin{equation}\label{non.extensive.th}
N\rightarrow+\infty:N\mathcal{A}=const\mbox{ and } \frac{U}{N^{\frac{7}{3}}}=const,
\end{equation}
where $\mathcal{A}=2\pi^{2}R^{3}$ is the hyper-surface area (or volume) of the sphere $\mathbb{S}^{3}$. This thermodynamic limit has been obtained in other astrophysical models \cite{Vel.QEM1}-\cite{Vel.QEM2}, which can be associated with the self-gravitating gas of non-relativistic identical points particles regardless their classical or quantum nature. The scaling $U\propto N^{7/3}$ also appears in Chandrasekhar theory of white dwarfs \cite{Chandrasekhar.31}-\cite{Chandrasekhar.35}. Even, the same scaling laws also appear in Thomas-Fermi theory \cite{Thomas}, where the asymptotic behavior of the atomic energy $E(Z)$ and density $\rho^{Z}(\mathbf{r})$ of an atom of charge $Z$ obey the following behaviors \cite{Scott}-\cite{Lieb2}
\begin{eqnarray}
E(Z)=C_{TF}Z^{7/3}+O(Z^{2})\mbox{ and }\rho^{Z}(\mathbf{r})\sim\rho^{Z}_{TF}(\mathbf{r})=Z^{2}\rho^{1}_{TF}\left(Z^{1/3}\mathbf{r}\right)
\end{eqnarray}
for $Z$ sufficiently large. As expected, the present scaling relations hold for point particles that interact by means of $1/r^{2}$ forces.

\begin{figure*}
\centering
\includegraphics[width=6.0in]{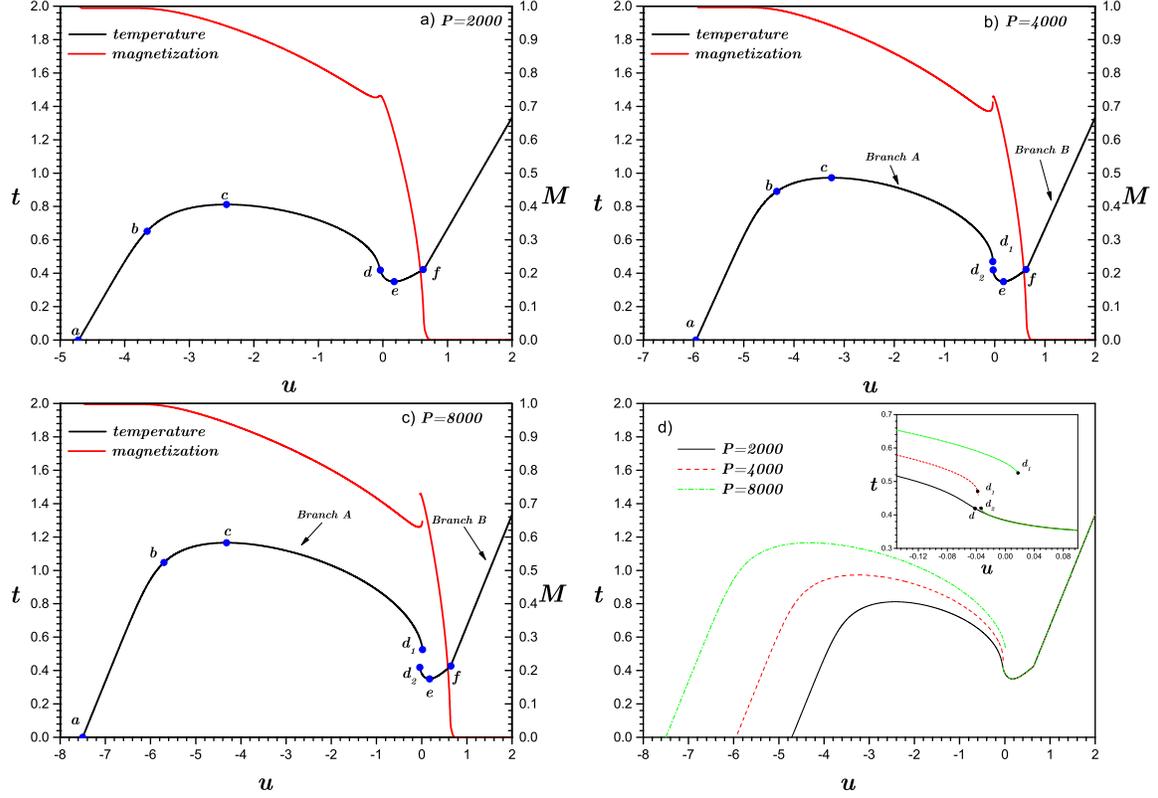}\\
\caption{Panels a)-c) Dependencies of the dimensionless temperature $t=kTR/GMm$ and magnetization $\mathbf{M}$ \emph{versus} the dimensionless energy $u=UR/GM^{2}$ for different values of partition number $P$ of the sphere $\mathbb{S}^{3}$. Panel d) Comparison among the caloric curves for different values of the partition number $P$. Inset panel: A zoom of these curves near the point of gravitational collapse $u_{d}$, which evidences the \emph{discontinuous} character of this microcanonical phase transition for values sufficiently large of the number of partitions $P$.}\label{todo.eps}
\end{figure*}

Let us now show that the relevance of the above thermodynamic limit is supported by a sort of an astrophysical counterpart of the known \emph{Gibbs-Duhem relation}. The temperature $T$ and the pressure $p$ are obtained from the thermodynamics relations:
\begin{equation}\label{RT}
\frac{1}{T}= \left(\frac{\partial S}{\partial U}\right)_{V} \mbox{ and }  \frac{p}{T} =\left( \frac{\partial S}{\partial V}\right)_{U}.
\end{equation}
For the present situation, the volume $V$ should be replaced by the hyper-surface area $\mathcal{A}$ of the sphere $\mathbb{S}^{3}$, while the pressure $p$ by surface tension $\sigma$. This procedure leads to the relations:
\begin{equation}\label{D}
\frac{1}{T}= \frac{Nk}{U_{c}}f^\prime(u) \mbox{ and }  \frac{\sigma}{T}  =  Nkf^{\prime }\left( u\right) U\frac{1}{3\mathcal{A}U_{c}} +\frac{1}{2}\frac{Nk}{\mathcal{A}}.
\end{equation}
Combining these results, one obtains:
\begin{equation}\label{estado}
\sigma\mathcal{A}=\frac{1}{2}NkT+\frac{1}{3}U.
\end{equation}
Considering the inverse temperature relation (\ref{cond.rest}):
\begin{equation}
U=\frac{3}{2}NkT+\left\langle W\right\rangle\equiv\left\langle K\right\rangle+\left\langle W\right\rangle,
\end{equation}
the state equation is expressed in terms of the average value of the  total potential gravitational energy $W$:
\begin{equation}\label{state.equation}
\sigma\mathcal{A}=NkT+\frac{1}{3}\left\langle W\right\rangle.
\end{equation}
An analogous result was obtained by de Vega and Sanchez in their analysis of Antonov isothermal model \cite{Vegab}-\cite{Vegac}. Notice that equation of state for the surface tension $\sigma$ is equivalent to the known virial theorem:
\begin{equation}\label{virial}
3\sigma\mathcal{A}=2\left\langle K\right\rangle+\left\langle W\right\rangle.
\end{equation}

\begin{figure*}
 \centering
  \includegraphics[width=6.0in]{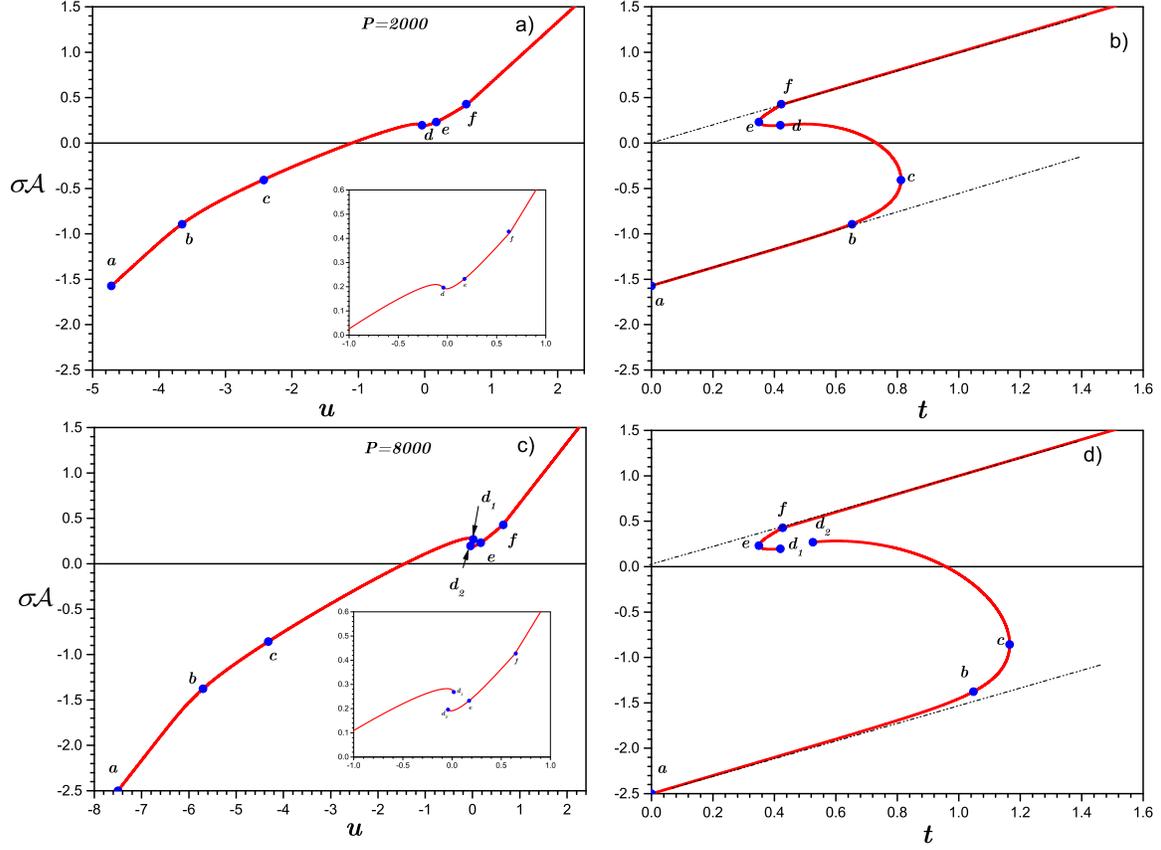}\\
  \caption{Equation of state for the present model: the surface tension times the hyper-surface area $\sigma \mathcal{A}$ for a number of partitions $P=2000$ and $8000$. Panels a) and c) this observable as a function on the dimensionless energy $u=UR/GM^{2}$. Panels b) and d) the same quantity as a function on the dimensionless temperature $t=kTR/GMm$.} \label{LEcuaEs.eps}
\end{figure*}

The chemical potential $\mu$ is obtained from the identity:
\begin{equation}\label{chem}
-\frac{\mu}{T}=\frac{\partial S}{\partial N},
\end{equation}
which yields:
\begin{equation}
-\frac{1}{T}\mu N=S+\frac{1}{2}Nk-2\frac{1}{T}U.
\end{equation}
Rephrasing the state equation (\ref{state.equation}) as follows:
\begin{equation}
\frac{1}{2}Nk=\frac{\sigma\mathcal{A}}{T}-\frac{1}{3}\frac{U}{T},
\end{equation}
one obtains a sort of Gibbs-Duhem relation for the present astrophysical situation:
\begin{equation}\label{Duhem-Gibbs}
S=\frac{1}{T}\left[\frac{7}{3}U-\sigma\mathcal{A}-\mu N\right].
\end{equation}
This result is straightforwardly obtained from the scaling behavior:
\begin{equation}
S(\alpha^{7/3}U,\alpha^{-1}\mathcal{A},\alpha N)=\alpha S(U,\mathcal{A},N),
\end{equation}
which is in fully agreement with the thermodynamic limit (\ref{non.extensive.th}). Astrophysical Gibbs-Duhem relation (\ref{Duhem-Gibbs}) was obtained in the precedent paper \cite{Vel.OpenAstro} for the case of Antonov isothermal model. This result is also related to the proposal of Latella et al. \cite{Latella.2015}, but its derivation obey to a different perspective: the scaling transformations of entropy that generalize the known extensive properties of conventional systems:
\begin{equation}\label{ext.ent}
S(\alpha U,\alpha V,\alpha N)=\alpha S(U, V, N)\rightarrow S=\frac{1}{T}\left[U+pV-\mu N\right].
\end{equation}
Identity (\ref{Duhem-Gibbs}) exhibits the same geometrical significance of its conventional expression (\ref{ext.ent}).

\begin{figure*}
  \centering
  \includegraphics[width=6.0in]{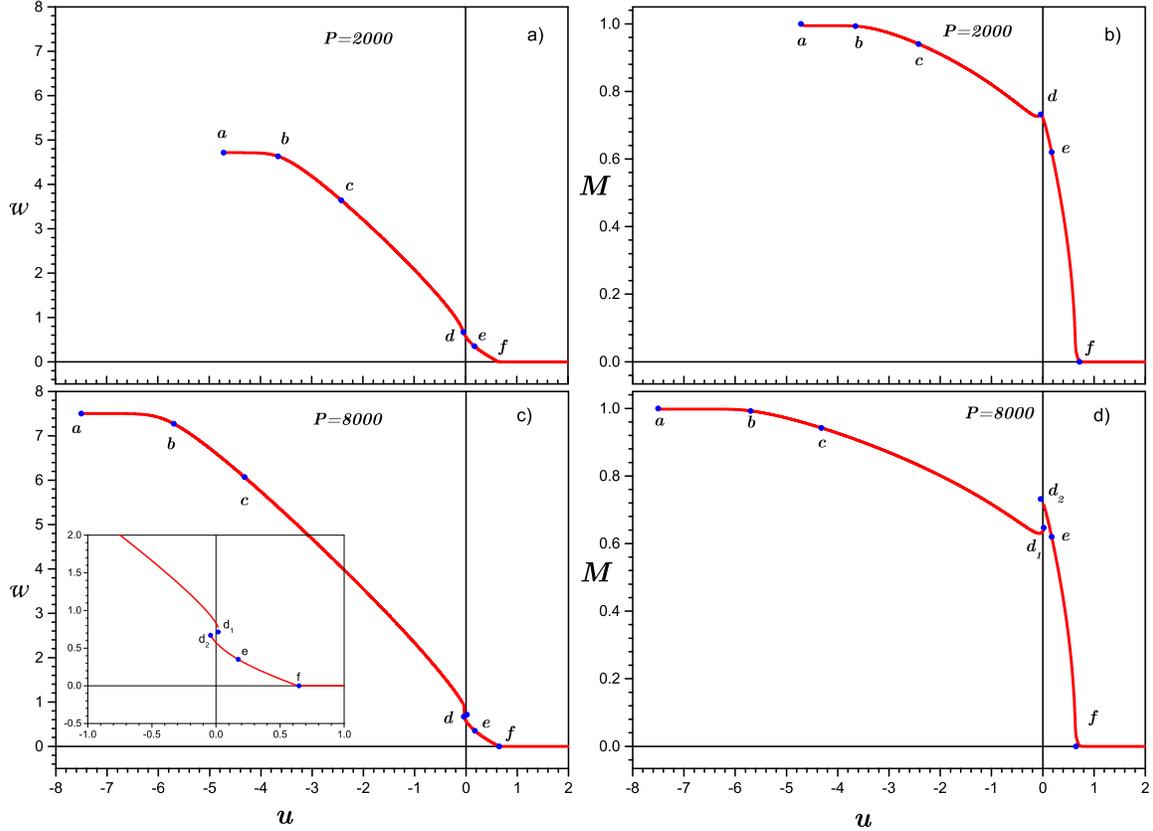}\\
  \caption{Behavior of order parameters $w=|W|R/GM^{2}$ and $M$ \emph{versus} the dimensionless energy $u=UR/GM^{2}$ for values of the partition number $P=2000$ and $8000$. Both order parameters have zero value for energies $u\geq u_{f}$. Once occurred the gravitational clustering, the breakdown of homogeneity in particles distribution explains the non-vanishing of these order parameters for $u<u_{f}$. During the occurrence of gravitational collapse, the absolute value of gravitational potential increases but the magnetization $M$ undergoes an abrupt fall. This behavior is related to a mass displacing towards the outer regions during the formation of a core-halo structure (see in figure \ref{profiles.eps}).}\label{orden.eps}
\end{figure*}

The presence of the length $\ell_{G}$ with a quantum-gravitational significance in the second term of entropy (\ref{Sreducida}) is anything but trivial. Its physical meaning can be understood by analyzing the applicability of classical description for the present self-gravitating gas. Denoting by $d\sim n^{-1/3}$ the characteristic separation among the particles for the gas with mean density $n$, the quantum uncertainty of the total kinetic energy $K$ must satisfy the following condition:
\begin{equation}
N\frac{1}{2m}\left(\frac{\hbar}{d}\right)^{2}\ll \left\langle K\right\rangle\sim -\frac{1}{2}\left\langle W\right\rangle\sim \frac{GM^{2}}{2R}.
\end{equation}
Taking into account the expressions $M=Nm$ and $n\sim N/R^{3}$, one obtains:
\begin{equation}\label{classical.limit}
RN^{1/3}\gg \ell_{G}.
\end{equation}
Accordingly, the requirement that the product $\mathcal{A}N$ remains constant in the thermodynamic limit $N\rightarrow +\infty$ guarantees the licitness of classical description.

\begin{figure*}
  \centering
  \includegraphics[width=6.0in]{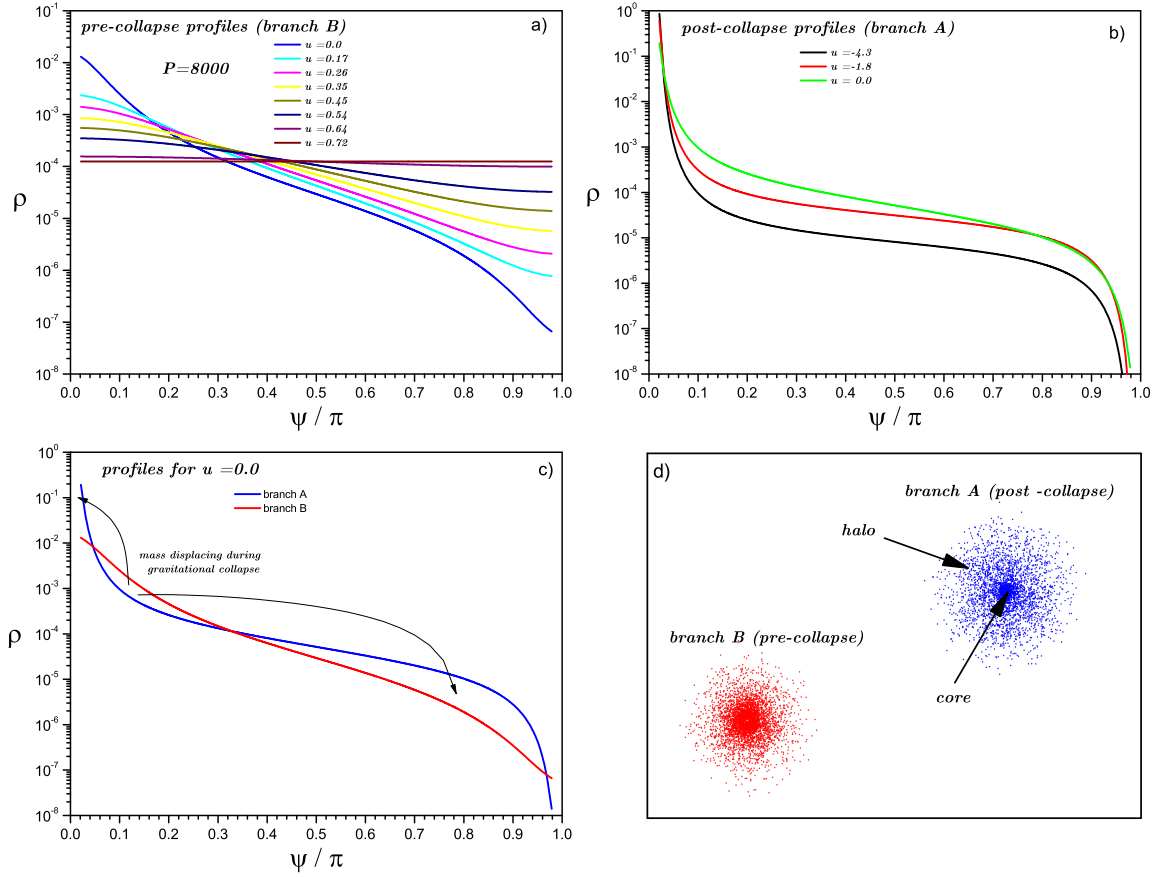}\\
  \caption{Distribution profiles for $P=8000$ and different values of dimensionless energy $u=UR/GM^{2}$. Panels a) and b) pre-collapse and post-collapse distribution profiles, respectively. Panels c) and d) Comparison of two coexisting distribution  profiles for the dimensionless energy $u=0.0$ and their schematic representations. Notice the occurrence of a mass displacing from the intermediate region towards inner and outer ones during the formation of a core-halo structure.}\label{profiles.eps}
\end{figure*}

\subsection{Macroscopic behaviors and clustering phase transitions}

Microcanonical dependencies of dimensionless temperature $t=kTR/GMm$ and magnetization $M$ for different values of the partition number $P$ are shown in figure \ref{todo.eps}. For the sake of convenience, the total energy $U$ is hereinafter replaced by the dimensionless energy $u=UR/GM^{2}$. The product of surface tension $\sigma$ and the surface area $\mathcal{A}$ described by the state equation (\ref{state.equation}) is shown in figure \ref{LEcuaEs.eps}. Actually, the quantity $\sigma\mathcal{A}$ was expressed into dimensionless units as $\sigma\mathcal{A}R/GM^{2}$, which is represented in terms of both the dimensionless temperature $t$ and the dimensionless energy $u$. On the other hand, microcanonical dependencies of order parameters $M$ and $w=|W|R/GM^{2}$ are shown in figure \ref{orden.eps}. Finally, dependence of particles distribution on different values of dimensionless energy $u$ for $P=8000$ are shown in figure \ref{profiles.eps}. According to these numerical results, our model system exhibits several notable points, and even, the existence of two separated branches in all its thermodynamic behaviors if the number of partitions $P$ is sufficiently large. Let us discuss these results with more details.

\begin{figure}
\centering
\includegraphics[width=4.0in]{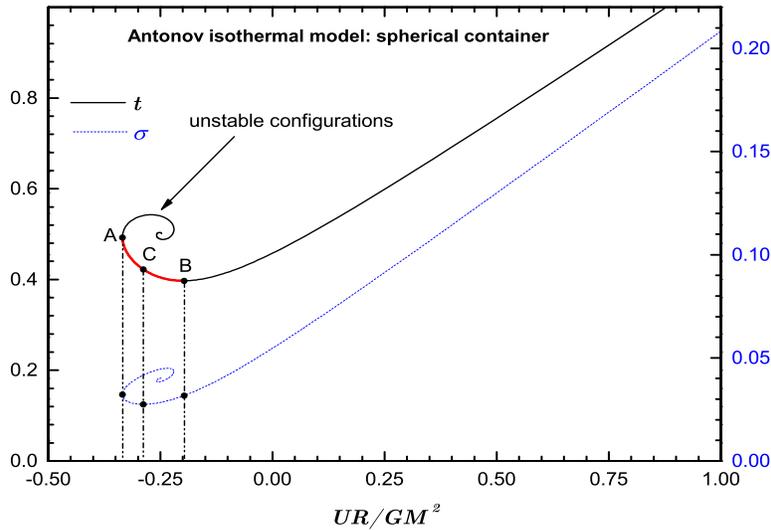}\\
\caption{Microcanonical temperature and pressure dependencies associated with Antonov isothermal model with spherical container using dimensionless variables $u=UR/GM^{2}$ (energy), $t=kTR/GMm$ (temperature) and $\sigma=pR^{4}/GM^{2}$ (pressure). The notable points of caloric curve $u_{A}=-0$.$355$ and $u_{B}=-0$.$197$ correspond to critical points of gravothermal and isothermal collapses. The branch $u_{A}<u<u_{B}$ (red line) are configurations with negative heat capacities. The microcanonical pressure exhibits a third notable point $u_{C}=-0$.$288$ where the dimensionless pressure variable $\sigma$ over exhibits its minimal value $\sigma_{C}=0$.$027$.}\label{Antonov.eps}
\end{figure}

The constituents of the self-gravitating gas are homogeneously distributed over the sphere $\mathbb{S}^{3}$ for energies $u>u_{f}\simeq0$.$623$. This system behaves as an ideal gas for these large energies, which is evidenced by the linear dependence between temperature and energy in figure \ref{todo.eps}, as well as the licitness of ideal gas equation of state $\sigma \mathcal{A}=NkT$ (see in figures \ref{LEcuaEs.eps}.c and \ref{LEcuaEs.eps}.d). As expected, the value of the magnetization $M$ and the total gravitational potential are zero for these energies. However, these order parameters exhibit an abrupt growth for energies $u\lessapprox u_{f}$, which reveal the nonhomogeneous character of particles distribution in this energy region. Accordingly, the notable point $u_{f}$ corresponds to the occurrence of a first clustering phase transition, when one observes the breakdown of $SO(4)$ symmetry of our model system. Formally, this collective behavior represents a \emph{continuous microcanonical phase transition}, which is quite similar to the one observed in the called HMF model \cite{Antoni,Chavanis2005,Konishi}. The existence of this collective behavior is a distinguishing feature of the present astrophysical thermo-statistical model. In fact, this phase transition is not observed in all those astrophysical models defined on a subset of the real space $\mathbb{R}^{3}$, which avoid the incidence of evaporation considering an external confining potential field. A particular example is the case of Antonov isothermal model \cite{Antonov}:
\begin{equation}\label{sgg}
H=\sum_{i}\frac{1}{2m}\mathbf{p}^{2}_{i}+w_{c}(\mathbf{x}_{i})-\sum_{i<j}\frac{Gm^{2}}{\left|\mathbf{x}_{i}-\mathbf{x}_{j}\right|},
\end{equation}
which is a self-gravitating gas of identical point particles that is enclosed into a subset $\Omega$ of the real space $\mathbb{R}^{3}$ using a container with impenetrable walls. Mathematically, this container is described here by the external potential field:
\begin{equation}
w_{c}(\mathbf{x})=\left\{
\begin{array}{cc}
0 & \mbox{if } \mathbf{x}\in \Omega\subset \mathds{R}^{3}\\
+\infty & \mbox{otherwise.} \\
\end{array}
\right.
\end{equation}
For comparison purposes, the caloric curve and the pressure of this second model are shown in figure \ref{Antonov.eps} when the container boundary $\delta \Omega$ is a sphere $\mathbb{S}^{2}\subset \mathbb{R}^{3}$ of radius $R$. This model system asymptotically behaves as an ideal gas for very large energies, which means that particles distribution is \emph{rigorously homogeneous} in the limit $u\rightarrow+\infty$ only. Hamiltonian (\ref{sgg}) of Antonov isothermal model does not obey the invariance under translations, and therefore, particles distribution is rigorously nonhomogeneous for any finite energy. In contrast, Hamiltonian (\ref{hamilton}) of our model system is invariant under $SO(4)$ rotations, which is the reason why the particles distribution can be homogeneous when the dimensionless energy $u>u_{f}$.

The incidence of gravitation turns more and more important when the total energy of the system decreases. The predominance of gravitation leads to the existence of states with negative heat capacities for energies $u_{d}<u<u_{e}$, where $u_{d}=-0$.$042$ and $u_{e}=0$.$173$. According to virial relation (\ref{virial}), states with negatives heat capacities can appear when the absolute values of surface tension $\sigma$ are sufficiently small. The notable point $u_{e}$ is a local minima of the microcanonical caloric curve, which corresponds to the critical point of called \emph{isothermal collapse}. In general, this gravitational instability is not relevant for real astrophysical situations since it requires the \emph{thermal contact} with a heat bath with infinite heat capacity. The usage of this argument is very usual in conventional extensive systems of everyday practice (systems with short-range interactions) that are under thermodynamic influence of the natural environment. However, astrophysical systems are driven by gravitation, which is a long-range interaction. In general, the consideration of a thermal contact with a heat bath is not a licit argument in presence of long-range interactions. The interaction among astrophysical systems always involves each one of these systems as a whole, which contrasts with the \emph{effective surface character} of the interactions associated with thermal contact in conventional applications \cite{Vel.OpenAstro}. Moreover, the canonical description (associated with the thermal contact of our system) is nonequivalent to microcanonical description discussed here. The microcanonical caloric curves of figure \ref{todo.eps} exhibit several values of dimensionless energy $u=UR/GM^{2}$ when the values of dimensionless temperature $t=kTR/GMm$ belongs to the interval $t_{e}\leq t \leq t_{c}$. Ensemble inequivalence arises here because of there is no one-to-one (bijective) correspondence between these thermodynamic variables. In fact, configurations with negative heat capacities that belong to the energy interval $u_{c}< u< u_{e}$ are non-accessible within canonical description.

According to dependencies shown in figure \ref{todo.eps}.d, the thermodynamic behavior of our model system is almost independent on the number of partitions $P$ for energies $u>u_{d}$. However, a very different situation is observed for lower energies. Microcanonical caloric curve for $P=2000$ exhibits a change of monotony at the notable energy $u_{d}$ rather similar to the one observed at the critical energy $u_{f}$, while the magnetization order parameter $M$ exhibits a local maximum (see in figure \ref{todo.eps}.a). On the contrary, one observes a \emph{discontinuous jump} in dimensionless temperature dependence for the case $P=8000$. Consequently, the notable point $u_{d}$ corresponds to the occurrence of a second clustering phase transition of our model system, whose character continuous or discontinuous crucially depends on the number of partitions $P$: it is \emph{continuous} when the values of $P$ are low enough (as the case $P=2000$), while it turns \emph{discontinuous} for $P$ sufficiently large (as the case $P=8000$). All these thermodynamic dependencies exhibit two stable branches that are denoted here by the letters $A$ (low energy behavior) and $B$ (high energy behavior).

This second microcanonical phase transition represents a special form of gravitational clustering: the called \emph{gravitational collapse}, where one observes the formation of a very \emph{dense core} in the inner regions of the system (see in figure \ref{profiles.eps}.d). This is the called \emph{gravothermal catastrophe} of the Antonov isothermal model \cite{Antonov}, which takes place at the notable point $u_{A}=-0.335$ as shown in figure \ref{Antonov.eps}. Gravothermal collapse is a thermodynamic instability that arises when the internal pressure of the self-gravitating gas (or its surface tension $\sigma$ in our case) is unable to compensate the incidence of its own gravitational forces. As expected, the stable branch $A$ with low energies describes the system thermodynamic behavior for post-collapse configurations.

\begin{figure*}
  \centering
  \includegraphics[width=6.0in]{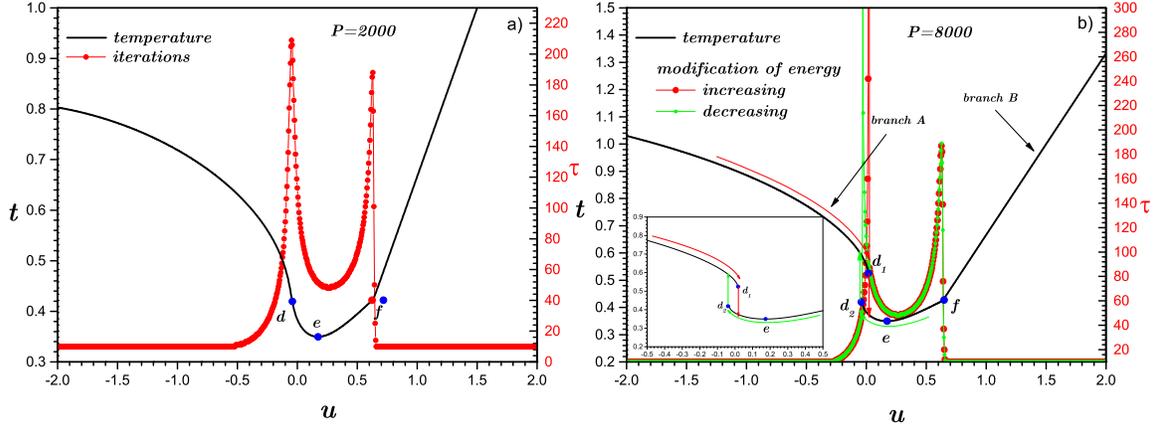}\\
  \caption{Other indicator about the occurrence of microcanonical phase transitions of this model is the energy dependence of the number of iterations $\tau$ that is necessary for the convergence of the scheme of successive iterations during numerical solution for the self-consistent problem (\ref{Phi.matrix})-(\ref{beta.disc}). We show here results for $P=2000$ and $8000$. Since the discontinuous jump associated with the gravitational collapse for $P=8000$, one can observe that the passage from one branch to another one does not occur at the same energy point when the energy is increased (forward direction) or decreased (backward direction) [see in the inset of panel b]. This sort of \emph{hysteresis phenomenon} arises from the presence of the meta-stables states.}\label{hysteresis.eps}
\end{figure*}

At first glance, dependence of the low energy thermodynamic behavior on the number of partitions $P$ can be regarded as an artificial result of our computational study, overall, since $P$ is merely a parameter of finite element methods employed for the computational study of the self-consistent problem (\ref{Phi.matrix})-(\ref{beta.disc}). However, the number of partitions $P$ formally defines the minimum volume $\delta \mu=\mathcal{A}/P$ to be occupied by the self-gravitating gas, a constraint that acts here as a \emph{regularization prescription} for the short-range divergence of the gravitational interaction (\ref{interact.w}) when particles separation $s\left(\mathbf{x}_{i},\mathbf{x}_{j}\right)\rightarrow0$. Previously, we have assumed that particles distribution are homogeneous inside each partition of the sphere $\mathbb{S}^{3}$, which is formally equivalent to impose that the particles that are located inside a same partition are \emph{non-interacting}. A more rigorous mathematical treatment to implement this prescription is to regularize the gravitational potential (\ref{interact.w}) by introducing a small cutoff length $a$ as follows:
\begin{equation}
w_{R}\left(\mathbf{x}_{i},\mathbf{x}_{j}\right)=m_{i}m_{j}\left\{
\begin{array}{cl}
  +\mathcal{G}(a) & \mbox{ if }0\leq s_{ij}\leq a, \\
  \mathcal{G}(s_{ij}) & \mbox{ if }a<s_{ij}<\pi R-a, \\
  -\mathcal{G}(a) & \mbox{ if }\pi R-a\leq s_{ij}\leq \pi R,
\end{array}
\right.
\end{equation}
where $s_{ij}\equiv s\left(\mathbf{x}_{i},\mathbf{x}_{j}\right)$ is the separation between the particles. Heuristically, the cutoff length $a$ and the number of partitions $P$ could be related as $a\propto \left(\mathcal{A}/P\right)^{1/3}$. This type of prescription enables the system to find a new branch of thermodynamic stability after the occurrence gravitational collapse. Notice that the post-collapse branch $A$ is not present in the microcanonical caloric curve of Antonov isothermal model shown in figure \ref{Antonov.eps}. In other words, the incidence of gravitation alone is unable to explain the existence of the post-collapse configurations, that is, additional physical considerations are required to describe interactions among the particles at very small distances. The argument considered in the present study is just a particular example. As already discussed by Chavanis \cite{Chavanis2002}, other astrophysical models include the incidence of quantum effects (e.g.: Pauli exclusion principle) or the consideration of hard-core (soft-core) particles with finite size. For example, it is usual in cosmological $N$-body simulations to employ the following short-range regularization of Newtonian interaction potential:
\begin{equation}\label{reg.pot}
V_{a}\left( \mathbf{r}_{1},\mathbf{r}_{2}\right) =-Gm_{1}m_{2}/\sqrt{\left|\mathbf{r}_{1}-\mathbf{r}_{2}\right|^{2}+a^{2}}
\end{equation}
between two particles with mass $(m_{1},m_{2})$ with a separation distance $r_{12}=\left|\mathbf{r}_{1}-\mathbf{r}_{2}\right|$. The length scale $a$ describes the existence of the \emph{internal structure} for each constituting particle (a soft core). For separation distances sufficiently large, this mathematical form of regularized potential can be obtained if each constituting particle obeys the density profile of the known Plummer model \cite{Plummer}:
\begin{equation}\label{Plumer.prof}
n(\mathbf{r})=\left(\frac{3m}{4\pi a^{3}}\right)\left(1+\frac{r^{2}}{a^{2}}\right)^{-5/2}.
\end{equation}
According to the present analysis, the collapsed structures resulting from these dynamical simulations necessarily depend on this type of short-range regularization.

According to dependencies shown in figure \ref{todo.eps}.d, there exists an energy interval where both stable branches are overlapped for $P=8000$. Specifically, one observes the existence of two different stable states for a same value of the dimensionless energy $u$ within the interval $u_{d_{2}}<u<u_{d_{1}}$. The end points of these branches have been labeled as $d_{1}$ (end point of branch B) and $d_{2}$ (end point of branch A). This coexistence of stable configurations evidences that the transition between the stable branches can occur in any energy of this interval. In fact, one should observe the occurrence of \emph{metastability}: a tendency of the system to remain trapped in a certain (metastable) configuration despite the other configuration exhibits a major occurrence probability (or a larger entropy). Imagine that our model system is firstly subjected to a process when its internal energy is increased, and afterwards, the same one is subjected to a second process where its internal energy is decreased. Since the transition from a stable branch to the other does not necessarily take place a the same energy point, one should observed a sort of \emph{hysteresis phenomenon}. This experiment was indeed developed in our computational study, whose results are reported in figure \ref{hysteresis.eps}. For a better understanding, we have included here the number of iterations $\tau$ considered to reach the convergence of calculations. As expected, the number of iteration exhibit a considerable increasing in the vicinity of the two microcanonical phase transitions. However, the slow relaxation is more severe for the case of discontinuous microcanonical phase transition for $P=8000$. Moreover, the transition from a branch to the other ones does not occur at the same energy point if simulations are performed increasing the total energy, and alternatively, by decreasing the energy.

The second clustering phase transition is located inside the energy region $u_{c}<u<u_{e}$ with negative heat capacities. The mere presence of these states reveals that our astrophysical system is nonhomogeneous. However, this behavior does not imply the coexistence of stable configurations for a same value of energy. As already emphasized, the coexistence of stable configurations (pre and post-collapse phases) is observed within the subset $u_{d_{2}}<u<u_{d_{1}}$ of this last interval when this transition is discontinuous. The great difference among configurations that belong from different branch reveals that the transition between these branches should be a violent process. According to distribution profiles shown in figures \ref{profiles.eps}.c and \ref{profiles.eps}.d, such a reconfiguration of particles distribution involves a great \emph{mass displacing} from the intermediate regions towards the inner and the outer regions, which leads to the formation of a very dense core and a more dense halo. This mass displacing explains the observed behavior of order parameters shown in figure \ref{orden.eps}. In particular, the mass displacing towards the outer regions explains the decreasing of the magnetization order parameter $M$, while the mass displacing towards the inner regions explains the growth of the absolute value of total potential energy $W$ (this behavior is better illustrated in the inset of figure \ref{orden.eps}.c).

The progressive reduction of the system energy provokes a continuous mass displacing from the halo towards the core (see in figure \ref{profiles.eps}.b). The occupation of the minimum volume practically occurs at point $u_{b}$, where the order parameter $M$ exhibits its maximum value close to the unity (see in Fig.\ref{orden.eps}). Simultaneously, the total potential energy $W$ reaches its minimum value $W_{min}\approx u_{a}GM^{2}/R$, which depends on the number if partitions $P$. However, dimensionless temperature $t_{b}$ at the notable energy $u_{b}$ is non-zero. As clearly evidenced in figure \ref{todo.eps} (as well as figures \ref{LEcuaEs.eps}.c and \ref{LEcuaEs.eps}.c), the self-gravitating gas recovers the ideal gas behavior for the energy interval $u_{a}<u<u_{b}$, that is, a linear dependence between temperature and energy, as well as the surface tension $\sigma$ and temperature. However, there is something different in this case: the extreme gravitational clustering of the self-gravitating gas produces \emph{negative values} in the surface tension $\sigma$ (see in figure \ref{LEcuaEs.eps}).

Thermodynamic behavior associated with the present model is in fully agreement with the qualitative picture discussed by Padmanabhan \cite{Padmanabhan1}. In particular, the long and short-range divergences of thermo-statistics of $N$-body gravitational problem in $\mathbb{R}^{3}$ were successfully avoided in our thermo-statistical study due to the consideration of long-range and short-range cutoffs $R$ and $a\propto\left(\mathcal{A}/P\right)^{1/3}$, which are respectively related to the radius of the sphere $\mathbb{S}^{3}$ and its partition into a finite number $P$ of cells. Moreover, the two clustering phase transitions reported here were anticipated by Klissing in a recent study for the case of sphere $\mathbb{S}^{2}\subset \mathbb{R}^{3}$ \cite{Kiessling}. Other related study is the called \emph{self-gravitating ring model} \cite{Sota,Tatekawa,Casetti}, which is 1D model defined on the sphere $\mathbb{S}\subset \mathbb{R}^{2}$. This toy model predicts behaviors quite similar to the model discussed by Padmanabhan, although it differs from the one discussed in this work for intermediate energies.

\section{Final remarks}

\subsection{Relevance in astrophysics and cosmology}

Despite its simplicity, the present model captures some aspects concerning to the formation of large astrophysical structures. The same one predicts that gravitational clustering can be associated with two microcanonical phase transitions. The formation of compact astrophysical structures is a collective behavior that is preceded by a gravitational clustering that involves a breakdown of homogeneity of particles distribution on the sphere $\mathbb{S}^{3}$. By itself, this picture suggests a hierarchical structuring: i) a first clustering phase transition drives the formation of distribution profiles with a larger linear dimensions, e.g.: a cloud or a nebulae, and ii) a second type of clustering phase transition, the gravitational collapse, leads to the formation of more compact structures with lower linear dimensions, that is, a very dense core (see in figure \ref{profiles.eps}.d). Moreover, this model describes that the formation of a very dense core during gravitational collapse is accompanied with a significant \emph{mass displacing} towards the outer regions (see in figure \ref{profiles.eps}.c) and the \emph{gravitational heating} of the whole astrophysical structure (see in figure \ref{todo.eps}). This mass displacing is indeed observed during gravitational collapse of a star, which is essentially non-equilibrium processes (due to the shock-wave that follows after the core-collapse). The present approach provides an alternative description to this process despite it explicitly invokes equilibrium, which evidences the power of the thermo-statistical description to deal with astrophysical situations.

Let us now refer to more subtle implications of our model, such as the proper influence of gravitational clustering on the equation of state of the matter in the Universe. According to relation (\ref{state.equation}), once the gravitational clustering breaks the homogeneity of matter distribution, a negative gravitational contribution to the pressure naturally appears. Thermodynamically speaking, this modification of the ideal gas equation of state $p=nkT$ is fully analogous to the one considered by Van der Waals equation of state \cite{Reichl}:
\begin{equation}\label{VdW}
p=\frac{nkT}{1-bn}-an^{2},
\end{equation}
where coefficients $(a,b)$ account for the incidence of attractive and repulsive intermolecular forces, respectively. While local inhomogeneities of matter distribution justifies the existence of this modification, the same one is systematically disregarded in most of approaches about the Universe expansion. For example, a typical assumption is that the equation of state that relates energy density $u$ and pressure $p$ in the stress-energy tensor for a perfect fluid \cite{Bergstrom}:
\begin{equation}
T_{\mu\nu}=(u+p)V_{\mu}V_{\nu}-pg_{\mu\nu}
\end{equation}
is fully determined by its proper microscopic nature in absence of gravitation, e.g.: the electromagnetic radiation obeys the relation $p=u/3$ or the non-relativistic degenerate gas $p=2u/3$. Apparently, the incidence of gravitation and their associated local heterogeneities seems to justify a modification of local surface tension $\sigma(\mathbf{x})$ of the self-gravitating gas as follows:
\begin{equation}\label{local.eos}
\sigma(\mathbf{x})=n(\mathbf{x})kT+\frac{1}{6}\rho(\mathbf{x})\Phi(\mathbf{x}).
\end{equation}
Here, $\Phi(\mathbf{x})$ is the gravitational potential, while $n(\mathbf{x})$ and $\rho(\mathbf{x})$ the densities of particles and mass, respectively. Note that expression (\ref{local.eos}) can be heuristically inferred from the equation of state (\ref{state.equation}). This result is obtained from (\ref{local.eos}) after integrating over the volume $d\mu(\mathbf{x})$ of the sphere $\mathbb{S}^{3}$ and replacing the local surface tension $\sigma(x)$ by its average value $\sigma$. Once gravitational collapse occurs, the gravitational contribution to the state equation turns dominant and explains the arising of negative pressures (or negative surface tension). Curiously, the existence of a fluid with a negative pressure is described by the $\Lambda$-term of Einstein field equation:
\begin{equation}
R_{\mu\nu}-\frac{1}{2}Rg_{\mu\nu}+\Lambda g_{\mu\nu}=\frac{8\pi G}{c^{4}}T_{\mu\nu},
\end{equation}
the called dark energy contribution \cite{Turner}. In the context of the evolutionary dynamics of the Universe, such a negative pressure explains an accelerated expansion. Although the influence of the local inhomogeneities over the large scale structure is usually neglected in cosmology \cite{Layzer2}, recent theoretical studies have suggested that they mimic effects usually attributed to cosmic acceleration from dark energy \cite{Wiltshire}. The present result is compatible with this type of arguments.

\subsection{Limitations and open problems}

A non-satisfactory aspect of the previous analysis is that gravitational clustering produces a single cluster, while a more realistic description should include the formation of multiple clusters and other more complex structures. Obvious limitations are the usage of steepest descend method and the simplifying assumption (\ref{azimuthal.sym}) considered in our computational procedure. A simple way to improve the present study is to replace the application of steepest descend and continuum approximation by a direct $N$-body Monte Carlo estimation of integral (\ref{NE1}). This type of approach enables the study of strong correlation effects, e.g.:, the calculation of two-point correlations function $\left\langle\delta\varrho(\mathbf{x})\delta\varrho(\mathbf{x}')\right\rangle$, which is of great interest in the study of large-scale distribution of matter \cite{Labini}.

The formation of multiple clusters and other structures could be favored by considering additional realistic ingredients. For example, the radius $R$ of sphere $\mathbb{S}^{3}$ is regarded here as a fixed parameter, while this quantity should be considered as a \emph{dynamical variable}. This type of peculiarity can be addressed by generalizing Hamiltonian model (\ref{hamilton}) as follows:
\begin{equation}\label{modif.H}
H=\frac{1}{2M}P^{2}+W(R)+\frac{1}{R^{2}}\sum_{i}\frac{1}{2m}\mathbf{p}^{2}_{i}+\frac{1}{R}\sum_{i<j}w\left(\mathbf{x}_{i},\mathbf{x}_{j}\right),
\end{equation}
where positions and momenta $\left(\mathbf{x},\mathbf{p}\right)$ are now referred to the unitary sphere $\mathbb{S}^{3}$. Moreover, the variable $P$ represents the momentum that is conjugated to the radius $R$, while the potential $W(R)$ can be employed to introduce other dynamical effects. This type of description introduces \emph{effective dynamical friction} in the microscopic dynamics of the self-gravitating gas. Other interesting ingredient is the \emph{coalescence} of system constituents. Certainly, the inclusion of these ingredients demands a more sophisticated thermo-statistical treatment, but they enable a theoretical analysis of some other interesting questions. For example, coalescence is the basic mechanism for the appearance of a mass spectrum for the constituents of the self-gravitating gas, while the incidence of effective friction associated with dynamical expansion of radius $R$ triggers the formation of \emph{dissipative structures} \cite{Philipson}. The above ingredients are actually considered in $N$-body cosmological simulations \cite{Konishi}, which are able to reproduce some features of large-scale distribution of galaxies reported by observations.

Other challenge for future works is the consideration of relativistic treatment of gravitation of the present thermo-statistics of $N$-body problem. As already shown, the microcanonical description requires consideration of the total energy and other integral of motions. However, the question of conservation laws involves some subtle difficulties in general relativity theory, such as inequivalence between their local and integral forms. A way to overcome this difficulty is the introduction of \emph{pseudo-tensors} for conserved quantities \cite{Landau}. While mathematical apparatus of general relativity presupposes an equivalent treatment for physical laws in all coordinate systems, one faces a different situation when conservation laws are described with pseudo-tensors: they leads to some mistaking artifacts of a particular coordinate system for real physical effects. Fortunately, energy conservation works out in the case of \emph{static spacetime} associated with an isolated $N$-body system in thermodynamic equilibrium. Other possible treatment is the foliation of spacetime described by \emph{ADM formalism} \cite{Arnowitt}, which enables the development of equations of motion for general relativity in the form of Hamilton's equations. In fact, these ideas have been exploited in the past to perform statistical-mechanic calculations in general relativity in the framework of microcanonical ensemble \cite{Brown}, specifically, for studies concerning to black hole thermodynamics. In the non-relativistic limit, the consideration of gravitational energy is crucial for conservation of total energy. Precisely, this feature is responsible at the macroscopic level of the gravitational modification to equation of state (\ref{state.equation}). It is quite interesting analyze to clarify whether or not this non-relativistic effect appears in general relativity.

\section*{Acknowledgements}

Authors thank to J. Ya\~{n}ez-Valenzuela and J.C. Rojas Gomez-Lobo for their suggestions and interesting discussions.

\end{document}